\documentclass[aps,prd,preprint,amssymb,eqsecnum,nofootinbib,floatfix]{revtex4}
\usepackage{epsfig}

\usepackage{amsmath}
\usepackage{amsfonts}
\usepackage{bm}
\usepackage{revsymb}
\usepackage{graphicx,epsfig}
\usepackage{placeins}
\newcommand{\be}{\begin{equation}}
\newcommand{\ee}{\end{equation}}
\newcommand{\bea}{\begin{eqnarray}}
\newcommand{\eea}{\end{eqnarray}}
\newcommand{\bes}{\begin{subequations}}
\newcommand{\ees}{\end{subequations}}

\begin{document}
\interfootnotelinepenalty=10000

% ----------------------------------------------------------------------
%
% TIME OF DAY
\newcount\hh
\newcount\mm
\mm=\time
\hh=\time
\divide\hh by 60
\divide\mm by 60
\multiply\mm by 60
\mm=-\mm
\advance\mm by \time
\def\hhmm{\number\hh:\ifnum\mm<10{}0\fi\number\mm}
\title{Modification to the Luminosity distance redshift relation in modified gravity theories}

% ----------------------------------------------------------------------

\author{\'Eanna \'E. Flanagan, Eran Rosenthal, and Ira M. Wasserman}
\affiliation{Center for Radiophysics and Space Research, Cornell
  University, Ithaca, New York, 14853}
\date{printed \today{} }

\begin{abstract}
We derive an expression for  the luminosity distance as 
a function of redshift for a flat Robertson-Walker spacetime perturbed by 
arbitrary scalar perturbations possibly produced by a modified gravity theory 
with two different scalar perturbation potentials. Measurements 
of the luminosity distance as function of redshift 
provide a constraint on a combination of the scalar potentials and so they 
 can complement weak lensing and other measurements in 
trying to distinguish among the various alternative theories of gravity.
\end{abstract}

\maketitle

\section{Introduction}

General relativity (GR) is in good agreement with all astrophysical observations 
of binary pulsars and solar system tests \cite{lrr-2001-4}. 
These observations provide tight constraints on deviations from GR  
on scales that are smaller or comparable with our solar system. 
Observations on larger scales are less restrictive, so it is possible that gravity is  
substantially different from GR on these scales. 
In recent years there has been a considerable effort to construct  theories 
 that modify GR on large scales, partly for the purpose of explaining 
the current phase of accelerated expansion of the Universe  
without introducing a dark energy component. 
 Among these theories are scalar tensor theories (see e.g. \cite{Perrottaetal,Boisseauetal}), 
f(R) theories \cite{carroll} and DGP gravity \cite{dgp}. Many studies have discussed comparing 
 observables in these theories with present and future observations, and in this way 
constraining  and sometimes refuting  these alternative theories of gravity 
(see e.g. \cite{Chiba,Wangetal,Ishaketal,Danieletal,zhangetal,zhangetal2}).
In this paper, we focus on one  observable -- the luminosity distance as function of redshift 
for a perturbed  Robertson-Walker (RW) Universe. 
We calculate  this observable for a  class of metric theories of gravity including GR.  
Measurements of the luminosity distance as function of redshift for type Ia supernovae (SNe)
provide evidence that the Universe expands at an accelerating rate \cite{perlmutteretal,riessetal}. 
These studies would be extended by the 
planned joint dark energy mission (JDEM). 
Observations of the luminosity distance as function of redshift 
may be able  to  constrain  the various alternatives to GR.

For any metric theory of gravity the RW metric provides a good description of the Universe on large scales.    
The construction of this metric is based on  observations that on large scales  
the Universe is homogeneous and isotropic (see e.g. \cite{wuetal,yadavetal}), 
and also on the assumption that the Copernican principle holds, 
namely that  we are not in any special location in the Universe.  
Since these general considerations are independent of the theory of gravity 
the construction of the RW metric remains valid for 
a large class of modified GR theories. 
Assuming the RW metric,  
one can normally adjust  the parameters of a given modified GR theory so that it would   
 produce the observed expansion history of the  Universe.  
However,  small deviations from homogeneity and isotropy 
which give rise to metric perturbations do depend on the particular theory of gravity
in  use. This makes the perturbed 
RW metric an ideal framework for 
studying observables that could distinguish among theories of gravity.

Many previous studies have suggested using cosmic microwave measurements, 
weak lensing measurements and other  observations  
 to distinguish among the various modified GR theories (see e.g. \cite{Wangetal,Ishaketal,Danieletal,zhangetal,zhangetal2}).  
Recently  Bonvin, Durrer, and Gasparini have  suggested that 
measurements of the  luminosity distance power spectrum induced by 
cosmological perturbations may be used to determine cosmological parameters \cite{bonvinetal}. 
Cooray,  Holz, and Huterer have showed that  two-point angular correlation function of SNe can 
provide useful data to study the foreground of large scale structure  \cite{coorayetal}.  
SNe surveys normally have smaller data sets then 
 weak lensing surveys so they typically have a larger statistical uncertainty.  Therefore,  
with regard to  sensitivity to cosmological fluctuations the  
SNe surveys are  not  competitive with weak lensing surveys. 
Nevertheless, they provide an 
independent  measurement of a different physical quantity (SNe surveys are sensitive to the luminosity of SNe 
while weak lensing surveys are sensitive to the distortion of galaxy images) 
 and so they can be used to  complement weak lensing measurements.

Another motivation to study the luminosity distance fluctuations  
is that they  degrade  the accuracy of the determination 
of cosmological parameters  from SNe data \cite{sarkaretal,huigreene,jonssonetal,coorayetalprl}.
Estimation  of the systematic error produced by cosmological perturbations 
 is therefore necessary for  SNe luminosity distance data analysis.

The lack of tight constraints on the 
theory of gravity on large scales 
together with the sensitivity of future SNe surveys to the  foreground 
of cosmological perturbations motivates the calculation of    
luminosity distance as function of redshift 
for a perturbed RW Universe in modified GR theories. 

In this paper we shall consider a 
flat RW metric with linear scalar perturbations. 
In a Newtonian gauge
this metric takes the form 
\begin{equation}\label{metric}
d{s}^2=a^2\biglb[-d\eta^2(1+2\psi)+(1-2\phi)\bold{dx}\cdot  \bold{dx}\bigrb]\,.
\end{equation}
Here $a(\eta)$ denotes the scale factor as function of conformal time, and 
the potentials $\phi$ and $\psi$ satisfy  $|\phi({\bf x},\eta)|,|\psi({\bf x},\eta)|<<1$. 
GR  in the absence of anisotropic stresses gives  $\psi=\phi$.
The luminosity distance for a perturbed RW Universe have been studied before 
by several authors. Thus Sasaki \cite{sasaki} has studied  the luminosity 
distance as function of redshift for a general perturbed spacetime. 
While Sasaki's analysis is very general, it gives 
an explicit expression for the luminosity distance 
only for the case  of an Einstein-deSitter Universe with $\psi=\phi$. For 
the case of $\phi=\psi$  an explicit expression 
for the luminosity distance was  derived by Pyne and  Birkinshaw  \cite {pynebirkinshaw} 
and was later corrected by  Hui and  Greene \cite{huigreene}. 
An equivalent  expression was recently derived by
Bonvin,  Durrer, and  Gasparini  \cite{bonvinetal}. 
In this paper we calculate the luminosity distance as function of redshift for the metric (\ref{metric}). 
We use a different method of calculation than the methods used in 
Refs. \cite{sasaki,pynebirkinshaw,huigreene,bonvinetal}, 
and our result generalizes the expressions in these references since we allow for 
 possibly different  scalar potentials $\phi\ne\psi$. 
Our  expression (\ref{finalresult}) reduces to  
the corresponding expression in Ref. \cite{huigreene}  
for the special case of $\phi=\psi$ [see Eq. (C21) in this reference].

Our complete expression for the luminosity distance ${D}_L$ as function 
of redshift for the metric (\ref{metric}) is given by a somewhat cumbersome formula (\ref{finalresult}).
 Fortunately, in practice most SNe surveys 
are sensitive only to subhorizon density perturbations. 
For these surveys, we may drop the terms that are subdominant for subhorizon perturbations,  
and obtain a simpler expression for the (subhorizon) luminosity distance ${D}^{sub}_L$ reading 
\begin{eqnarray}\label{finalapp}
&&{D}_L^{sub} ({z},\bold{n})  \approx(\chi_s-\chi_o)(1+{z})
\Biglb\{1+\bold{v}_s\cdot\bold{n}
-\frac{ (\bold{v}\cdot\bold{n})^{\chi_s}_{\chi_o}  }{(\chi_s-\chi_o)\mathcal{H}_s}
  \\\nonumber
&&-\frac{1}{2}\int_{\chi_o}^{\chi_s} \nabla^2(\phi+\psi)   
\frac{(\chi-\chi_o)(\chi_s-\chi)}{\chi_s-\chi_o}d\chi 
\Bigrb\}\,.
\end{eqnarray}
Here the luminosity distance is expressed in terms of the observed redshift ${z}$, and
the direction to the source, where $\bf{n}$  (also denoted as $n^a$) denotes 
 a unit spatial vector from the observer to the source.
The notation $\approx$ denotes an approximate equality accurate 
up to first-order in the potentials $\phi$,  $\psi$, and the peculiar velocities (in the conformal spacetime) 
$\bold{v}$.  The subscripts $s$ and $o$ refer to the 
source and the observer, respectively. 
The conformal Hubble rate is denoted $\mathcal{H}=\frac{da}{d\eta}a^{-1}$.
The potentials $\phi$ and $\psi$, which by definition are 
functions of the spacetime coordinates,  
are considered here to be functions of an affine parameter $\chi$ defined on 
the  zeroth order photon null geodesics. 
The affine parameter ${\chi}$ is an implicit function of the observed redshift ${z}$. This 
function is  determined by the unperturbed RW background, and is given by
\[
{\chi}_s=\int_0^{z} \frac{1}{H({z})}d{z}+{\chi}_o\,,
\]
where $H({z})\equiv \mathcal{H}/a$ is the Hubble rate,  $\chi_o$ denotes 
an arbitrary initial value for the affine parameter, and  we assume that the background spacetime
is either expanding or contracting.

We now briefly discuss  Eq. (\ref{finalapp}) and how it may be used to 
distinguish  among various theories of gravity. 
Notice that this formula has two types of term: terms that are proportional to the peculiar 
velocities of the observer and the source, these terms represent kinematical Doppler-shift, 
and a term depending on the Laplacian of the potentials that represents gravitational lensing. 
This lensing term depends only on the combination $\psi + \phi$ of the two potentials,
and so it can not be used to differentiate the GR case $\psi = \phi$ from the more general
case $\phi \ne \psi$. 
Fortunately, the velocity terms in Eq. (\ref{finalapp}) provide 
additional information that breaks this degeneracy. 
The dependence of the peculiar velocities on the potentials follows from the 
equations of motion of the observer and the source. We assume that the observer and the source 
feel no interaction other then gravity, and furthermore we assume 
that the total energy momentum tensor is  covariantly conserved. It now follows that 
the observer and the source move along geodesics of the perturbed spacetime, and therefore their peculiar velocities 
satisfy 
\begin{equation}\label{pecvel}
{\bf v}_{,\eta}+a_{,\eta}a^{-1}{\bf v}+\nabla\psi\approx 0\ .
\end{equation}
Since the peculiar velocities depend on $\psi$, but are independent of $\phi$, they provide information that breaks the degeneracy. 
It is also useful to examine the dependence on the redshift of the various 
terms in Eq. (\ref{finalapp}). 
Terms that depend on peculiar velocities are expected to be bounded 
 (the  peculiar velocity of a host galaxy is normally of order $500 {\rm km\ sec^{-1}}$) and therefore the
ratio of a typical peculiar velocity to the Hubble flow is larger at low redshift and smaller at high redshift.
 We might also expect the RMS effect due to lensing to increase with redshift. Therefore,  
 the lensing term is expected to be dominant at high redshift, while 
 the peculiar velocity terms are expected to become dominant at low redshift.  

Information about the scalar perturbations can be extracted from the data by 
calculating the  correlation function  
$[\bar{D}_L (z')\bar{D}_L (z)]^{-1}\langle {D}_L ({z'},\bold{n'}) {D}_L ({z},\bold{n})\rangle$, where 
$\bar{D}_L (z)$ denotes the average over angles, and $\langle ... \rangle$ denotes an ensemble average. 
We anticipate that if we specialize to either low or high redshifts, this correlation function
would be sensitive to either the peculiar velocity terms or the lensing term. This means that by combining information from both 
low and high redshifts one can overcome the above mentioned degeneracy. Such 
measurements may be able to provide information that could distinguish among 
various theories of gravity.

To calculate the correlation function from theory, 
one must have some knowledge about the 
underlying theory of gravity. In particular one must know the relation between the 
 overdensity and the gravitational potentials $\phi$ and $\psi$.
Instead of specializing to a particular modified theory of gravity
it is possible to employ a parameterized 
framework  where the relation between the two potentials and 
the relations among the potentials and the overdensity are parameterized, such that  
 various modified gravity theories produce different values for the parameters (see e.g. Ref. \cite{Aminetal,Husawicki}). 
In such a framework one can use Eq.  (\ref{finalapp}) 
in combination with the power spectrum of density fluctuations
to obtain specific predictions for the power spectrum of the 
luminosity distance, which can then be compared with observations. 
In this paper we keep our assumptions about the theory of gravity at minimum,
and we defer the more detailed parameterized analysis to future work. Nevertheless, we can still  
make a rough order of magnitude estimate of the correlation function. 
It reasonable to expect that even in a modified gravity theory the potentials 
$\psi$ and $\phi$ while not precisely equal, should be of the 
same order of magnitude. This means that 
while the correlation function should differ 
from the one in GR, it is likely to be of the same 
order of magnitude.  
Ref. \cite{bonvinetal} shows that for GR and a 
CDM Universe the contribution to the correlation 
function form lensing alone can be as large as $10^{-5}$ for $z=z'=2$ at $l\approx 300$.
Ref. \cite{coorayetal} estimates that this lensing contribution to the cross-correlation function could
be detected with a signal to noise ratio of 10 with a survey of 10,000 SNe over $10 {\rm deg}^2$ between
redshifts of 0.1 and 1.7. 
While the signature of  modified gravity is likely to  have a smaller signal to noise, 
it may still be detectable in future SNe surveys.

This paper is organized as follows. 
In Sec. \ref{generalst} we present the framework for calculating luminosity distance 
in a general spacetime, in Sec. \ref{conformalang} we specialize to the metric (\ref{metric}) and  
simplify the calculation by introducing 
a transformation to a 
conformal spacetime,
 in Sec. \ref{initcondition} we impose initial conditions, and finally in 
Sec. \ref{results} we obtain our final expression (\ref{finalresult}).

\section{The Luminosity distance in a general spacetime}\label{generalst}

There is a well known general relation \cite{Etherington,Ellis} between the observed luminosity distance
${D}_L({z})$ and the observed angular diameter distance ${D}_A({z})$ which reads
\begin{equation}\label{dlda}
{D}_L({z})=(1+{z})^2 {D}_A({z})\,.
\end{equation}
This relation  is valid in  any metric theory 
provided that the linear momentum of photons is conserved, 
and so in particular it holds for the metric (\ref{metric}).  
Below we calculate $ {D}_A({z})$ which is a purely geometrical quantity, and substitute our result back into  
Eq. (\ref{dlda}) to obtain the luminosity distance ${D}_L({z})$. 

The angular diameter distance  ${D}_A({z})$ is defined in the following manner. 
Suppose that an observer views a sizable distant object (e.g. a distant galaxy or a structure of the CMB anisotropy) 
that subtends a small solid angle $\Delta{\Omega}$.     
Using  the geometric optics approximation we  can describe the electromagnetic radiation using 
a congruence of null geodesics. We consider geodesics 
that emanate from a vertex at the observer and propagate backwards in time towards the source.
The angular diameter distance ${D}_A$  is given by  
\begin{equation}\label{da}
{D}_A({\lambda})\equiv\sqrt{ \frac{\Delta{A}({\lambda})} {\Delta{\Omega}}} \,.
\end{equation}
Here ${\lambda}$ denotes the affine parameter along the congruence, and  ${\Delta}{A}({\lambda})$
denotes the transverse cross sectional area of the congruence at a fixed ${\lambda}$.

We pick a representative null geodesic from the thin congruence 
and denote its worldline with ${x}^\alpha({\lambda})$, and denote its  tangent null vector field 
with ${k}^{\alpha}=\frac{dx^\alpha}{d{\lambda}}$, where throughout Greek indices run from 0 to 3. 
In general the evolution of a thin null congruence is completely described by an expansion parameter ${\theta}$, a shear tensor 
${\sigma}_{\alpha\beta}$, and a rotation tensor  ${\omega}_{\alpha\beta}$, where the 
 expansion parameter takes the form of 
\begin{equation}\label{areaexpan}
{\theta}=\frac{1}{\Delta{A}} \frac{d}{d{\lambda}}\Delta{A}\,.
\end{equation}
We assume that  the rotation tensor  ${\omega}_{\alpha\beta}$ vanishes at the 
 observer, since this tensor satisfies a homogeneous transport equation (see e.g. \cite{Wald}); 
this initial condition forces it to vanish everywhere.  
Therefore,   Raychaudhuri's equation for the null congruence reads 
\begin{equation}\label{raych}
\frac{d{\theta}}{d{\lambda}}=-\frac{1}{2}{\theta}^2-{R}_{\alpha\beta}{k}^\alpha{k}^\beta\,
-{\sigma}_{\alpha\beta}{\sigma}^{\alpha\beta}\,.
\end{equation}
Eqs. (\ref{da},\ref{areaexpan},\ref{raych}) gives the focusing equation (see e.g. \cite{Peebles})
\begin{equation}\label{diffeq}
\frac{1}{{D}_A} \frac{d^2 {D}_A}{d{\lambda}^2}=-\frac{1}{2}({R}_{\alpha\beta}{k}^\alpha{k}^\beta 
+{\sigma}_{\alpha\beta}{\sigma}^{\alpha\beta})\,,
\end{equation}
which we solve to obtain  ${D}_A$.

\section{Conformal angular diameter distance}\label{conformalang}

We now specialize the calculation of  ${D}_A$ to the metric (\ref{metric}). 
This metric allows us to factor out the dependence of  ${D}_A$  on the scale factor, and thereby 
simplify the calculation. To this end we consider the following  conformal transformation  
\begin{equation}\label{conformal}
d{s}^2=a^2 d\tilde{s}^2 \ \ ,\ \ \delta{\Omega}=\delta\tilde{\Omega}\ \ ,\ \ {D}_A=a\tilde{D}_A \ \ , 
d{\lambda}=a^2d\tilde{\lambda}\ \ , \ \ {k}^\alpha=a^{-2}\tilde{k}^\alpha\,.
\end{equation}
Here quantities with and without tildes denote  the conformal space and 
the real space, respectively. 
Notice that the form of Eq. (\ref{diffeq}) is invariant under this conformal transformation, this means that the focusing equation  
  in the conformal spacetime is obtained by adding tildes to all the quantities in Eq. (\ref{diffeq}). 
The focusing  equation in the conformal spacetime  can be 
transformed into an integral equation of the form
\begin{equation}\label{inteq}
\tilde{D}_A=-\frac{1}{2}\int_{\tilde{\lambda}_o}^{\tilde{\lambda}_s}
(\tilde{R}_{\alpha\beta}\tilde{k}^\alpha \tilde{k}^\beta +{\tilde{\sigma}}_{\alpha\beta}{\tilde{\sigma}}^{\alpha\beta})
\tilde{D}_A(\tilde{\lambda})(\tilde{\lambda}_s-\tilde{\lambda})d\tilde{\lambda}+\tilde{C}_1+\tilde{C}_2\tilde{\lambda}_s\,.
\end{equation}
Here $\tilde{C}_1$ and $\tilde{C}_2$ are constants which depend on the initial conditions, and all the quantities in the first
brackets on the right hand side are evaluated at $\tilde{\lambda}$.
So far we have not  used a perturbative 
approximation and Eq. (\ref{inteq}) is accurate to all orders 
in the perturbation potentials.
Below we calculate $\tilde{D}_A$ to the first order in the potentials $\phi$ and $\psi$.  

Expanding  the null geodesic worldline $x^\alpha(\tilde{\lambda})$  in a perturbation series 
gives 
\begin{equation}\label{pert1}
{x}^\alpha(\tilde{\lambda})\approx\bar{x}^\alpha(\tilde{\lambda})+\delta x^\alpha(\tilde{\lambda})\ ,
\ \tilde{k}^\alpha(\tilde{\lambda})\approx\bar{k}^\alpha(\tilde{\lambda})+\delta \tilde{k}^\alpha(\tilde{\lambda})\,,
\end{equation}
where an overbar denotes an unperturbed quantity, and a $\delta$ preceding a quantity 
denotes the first-order perturbation to that quantity. 
In this perturbation scheme
the  affine parameter  $\tilde{\lambda}_s$ 
characterizes the location of the source on the null geodesic. 
The same $\tilde{\lambda}_s$ is used both for  
 the perturbed quantity and for the unperturbed quantity. 
Notice, however, that due to the perturbation potentials 
the source is characterized by different coordinates  in the background spacetime 
and in the full spacetime $x^\alpha(\tilde{\lambda}_s)\ne\bar{x}^\alpha(\tilde{\lambda}_s)$.
Using the above notation we also have
\begin{equation}\label{pert2}
\tilde{D}_A\approx\bar{D}_A+\delta \tilde{D}_A\ , \ \tilde{R}_{\alpha\beta}\approx\delta \tilde{R}_{\alpha\beta}\ , 
\ \ \tilde{\sigma}_{\alpha\beta}\approx\delta\tilde{\sigma}_{\alpha\beta}
\ ,\ \tilde{C}_{1,2}\approx\bar{C}_{1,2}+\delta \tilde{C}_{1,2}\,.
\end{equation}
Notice that at the leading order the conformal spacetime is flat so that  the expansions for $\tilde{R}_{\alpha\beta}$ and 
$\tilde{\sigma}_{\alpha\beta}$ start at first-order.
Using Eq.(\ref{inteq}) together with Eqs. (\ref{pert1},\ref{pert2}) we obtain
\begin{eqnarray}\label{bard}
&&\bar{D}_A(\tilde{\lambda}_s)=\bar{C}_1+\bar{C}_2\tilde{\lambda}_s\,,\\ \label{deltad}
&&\delta \tilde{D}_A(\tilde{\lambda}_s)=\delta \tilde{C}_1+\delta \tilde{C}_2 \tilde{\lambda}_s 
-\frac{1}{2}\int_{\tilde{\lambda}_o}^{\tilde{\lambda}_s}
\delta \tilde{R}_{\alpha\beta}\bar{k}^\alpha \bar{k}^\beta
\bar{D}_A(\tilde{\lambda})(\tilde{\lambda}_s-\tilde{\lambda})d\tilde{\lambda}\,.
\end{eqnarray}
The flatness of the background spacetime implies that the background 
 null vector $\bar{k}^\mu$ is a constant four-vector. By rescaling $\tilde{\lambda}$  we set this vector to be  
\[
 \bar{k}^\mu=\frac{d\bar{x}^\mu}{d\tilde{\lambda}}=(-1,\bold{n})\,,
\]
 where $\bold{n}$ is a unit vector by virtue of the nullity of $\bar{k}^\mu$.
Locating  the observer at the origin $\bar{x}^a_o=0$, where $a,b=1,2,3$, we  have
\begin{equation}\label{flatsol}
\bar{\eta}(\tilde{\lambda})=\eta_o-(\tilde{\lambda}-\tilde{\lambda_o})\ ,\ \bar{x}^a(\tilde{\lambda})=n^a(\tilde{\lambda}-\tilde{\lambda}_o)\,.
\end{equation}

\section{Initial conditions}\label{initcondition}

We now  supplement equations (\ref{bard},\ref{deltad}) with  initial conditions at $\tilde{\lambda}=\tilde{\lambda}_o$, and 
thereby determine the constants $\tilde{C}_1$ and $\tilde{C}_2$. 
First we demand that 
  $\tilde{D}_A(\tilde{\lambda}_o)=0$, which gives 
\[
\bar{C}_1=-\tilde{\lambda}_o\bar{C}_2\ ,\ \delta\tilde{C}_1=-\tilde{\lambda}_o\delta\tilde{C}_2\,.
\]
We therefore have at the leading order  $\bar{D}_A(\tilde{\lambda}_s)=\bar{C}_2(\tilde{\lambda}_s-\tilde{\lambda}_o)$. 
Eq. (\ref{flatsol}) implies that  $\bar{C}_2=1$. With the above initial condition 
 Eqs. (\ref{bard},\ref{deltad}) read
\begin{eqnarray}\label{bard2}
&&\bar{D}_A(\tilde{\lambda}_s)=\tilde{\lambda}_s-\tilde{\lambda}_o\\ \label{deltad2}
&&\delta\tilde{D}_A(\tilde{\lambda}_s)=\delta \tilde{C}_2 \bar{D}_A(\tilde{\lambda}_s) 
-\frac{1}{2}\int_{\tilde{\lambda}_o}^{\tilde{\lambda}_s}
\delta \tilde{R}_{\alpha\beta}\bar{k}^\alpha \bar{k}^\beta
\bar{D}_A(\tilde{\lambda})(\tilde{\lambda}_s-\tilde{\lambda})d\tilde{\lambda}
\end{eqnarray}
The constant $\delta \tilde{C}_2$ is determined from the value 
of the derivative $\frac{d\tilde{D}_A}{d\tilde{\lambda}}$ at $\tilde{\lambda}_o$. 
To determine this constant it is instructive to  first  calculate 
 $\frac{d\hat{D}_A}{d\tilde{\lambda}}$,  where  $\hat{D}_A$ is 
the conformal angular diameter distance for a  comoving observer, ignoring 
for the moment the observer's  peculiar velocity. It is then possible 
 to  correct the expression and  account for the observer's  peculiar velocity. 
In the vicinity of a comoving observer we can Taylor expand the potentials 
$\phi$ and $\psi$ by treating the distance from the observer as the small parameter. 
It follows from the metric (\ref{metric}) that the angular diameter distance takes the form of 
\[
\hat{D}_A(\tilde{\lambda})=R (\tilde{\lambda})\sqrt{1-2\phi(\tilde{\lambda})}+O(R^{3/2})\,.
\]
where $R=(\delta_{ab}x^a x^b)^{1/2}$. 
Imposing initial conditions $\delta x^a(\tilde{\lambda}_o)=\frac{d\delta x^a}{d\tilde{\lambda}}(\tilde{\lambda}_o)=0$
gives 
\begin{equation}\label{dhatda}
\frac{d\hat{D}_A}{d\tilde{\lambda}}(\tilde{\lambda}_o)=1-\phi_o\,,
\end{equation}
which gives rise to a constant $\delta\hat{C}_2=-\phi_o$ for a comoving observer. 
To correct for the  observer's  peculiar velocity let us 
 consider  a transformation to a realistic reference   frame moving in 
a velocity  $\bold{v_o}$  with respect to the  
static observer in the conformal space, where $v_o=|\bold{v_o}|=O(\psi)=O(\phi)$. We maintain the notation 
$\tilde{D}_A$  for the angular diameter distance in the realistic frame, which  is given by 
\begin{equation}\label{datrans}
\tilde{D}_A(\tilde{\lambda})\equiv\sqrt{\frac{\Delta \tilde{A}}{\Delta \tilde{\Omega}}}=\sqrt{\frac{\Delta \hat{A}[1+O(v_o^2)]}{\Delta \hat{\Omega}
[1-2\bold{v_o} \cdot\bold{n}+O(v_o^2)]}}=\hat{D}_A(\tilde{\lambda})[1+\bold{v_o} \cdot\bold{n}+O(v_o^2)]\,.
\end{equation}
Using Eq. (\ref{deltad2}) together with Eqs. (\ref{dhatda},\ref{datrans})  we obtain
\[
\delta \tilde{D}_A(\tilde{\lambda}_s)= \bar{D}_A(\tilde{\lambda}_s)(\bold{v_o}\cdot\bold{n}-\phi_o) 
-\frac{1}{2}\int_{\tilde{\lambda}_o}^{\tilde{\lambda}_s}
\delta \tilde{R}_{\alpha\beta}\bar{k}^\alpha \bar{k}^\beta
\bar{D}_A(\tilde{\lambda})(\tilde{\lambda}_s-\tilde{\lambda})d\tilde{\lambda}\,.
\]
Evaluating the expression inside the integral gives
\begin{eqnarray}
&&\delta \tilde{D}_A(\tilde{\lambda}_s)= \bar{D}_A(\tilde{\lambda}_s)(\bold{v_o}\cdot\bold{n}-\phi_o)\\\nonumber 
&&-\frac{1}{2}\int_{\tilde{\lambda}_o}^{\tilde{\lambda}_s}
\left[ \nabla^2(\phi+\psi) +2\phi_{,\eta\eta} -4\bold{\nabla}\phi_{,\eta}\cdot\bold{n}+(\phi_{,ab}-\psi_{,ab})n^a n^b  \right]
\bar{D}_A(\tilde{\lambda})(\tilde{\lambda}_s-\tilde{\lambda})d\tilde{\lambda}\,.
\end{eqnarray}
Using the relation $\frac{d\phi}{d\tilde{\lambda}}=\nabla\phi\cdot\bold{n}-\phi_{,\eta}$ we integrate by parts 
and obtain
\begin{eqnarray}\label{deltad3}
&&\delta \tilde{D}_A(\tilde{\lambda}_s)= \bar{D}_A(\tilde{\lambda}_s)(\bold{v_o}\cdot\bold{n}-\phi_o)\\\nonumber 
&&-\frac{1}{2}\int_{\tilde{\lambda}_o}^{\tilde{\lambda}_s}
\left[ \nabla^2(\phi+\psi) -2\phi_{,\eta\eta} +(\phi_{,ab}-\psi_{,ab})n^a n^b  \right]
\bar{D}_A(\tilde{\lambda})(\tilde{\lambda}_s-\tilde{\lambda})d\tilde{\lambda}   
-2\int_{\tilde{\lambda}_o}^{\tilde{\lambda}_s}\phi_{,\eta}(\tilde{\lambda}_s+\tilde{\lambda}_o-2\tilde{\lambda})d\tilde{\lambda}\,.
\end{eqnarray}

\section{Redshift}

So far we have calculated the dependence of $\bar{D}_A$ and $\delta \tilde{D}_A$ on the affine parameter $\tilde{\lambda}$.  
The goal of  this section is to use these relations to express the  
 luminosity distance as function of the observed redshift ${z}$. 
Using  Eq. (\ref{pert2}) and Eq. (\ref{conformal}) together with Eq. (\ref{dlda}) we find that
\begin{equation}\label{dl1}
{D}_L(\tilde{\lambda}_s)\approx(1+{z})^2  a(\tilde{\lambda}_s)[\bar{D}_A(\tilde{\lambda}_s)+\delta \tilde{D}_A(\tilde{\lambda}_s)]\,.
\end{equation}
By definition the  observed redshift is given by
\begin{equation}\label{redshift}
1+{z}=\frac{{({g}_{\mu\nu} {k}^\mu {u}^\nu)}_{source}} 
{{({g}_{\alpha\beta} {k}^\alpha {u}^\beta)}_{observer}}\,.
\end{equation}
We also introduce  a conformal redshift 
$1+\tilde{z}$, defined  by adding tildes to all the quantities 
in Eq. (\ref{redshift}). 
Using Eqs. (\ref{redshift},\ref{conformal}) together with ${u}^\alpha=a^{-1} \tilde{u} ^\alpha$ we find that 
\begin{equation}\label{confredsh}
1+{z}=\frac{1+\tilde{z}} {a}\,,
\end{equation}
where for convenience we set $a(\tilde{\lambda}_o)=1$ at the observer and use the implicit 
notation $a\equiv a(\tilde{\lambda}_s)$.
Substituting Eq.(\ref{confredsh}) into Eq. (\ref{dl1}) gives
\begin{equation}\label{dl2}
{D}_L(\tilde{\lambda}_s)\approx(1+{z}) [(1+\tilde{z})\bar{D}_A(\tilde{\lambda}_s)+\delta \tilde{D}_A(\tilde{\lambda}_s)]\,.
\end{equation}
To obtain the deviation of the luminosity distance as function of redshift
due to the metric perturbations 
we need to compare the perturbed luminosity distance and the 
background luminosity distance at the same redshift.
Following Hui and Greene \cite{huigreene}
 we  calculate ${D}_L(\tilde{\lambda}_s+\delta\tilde{\lambda})$ 
the luminosity distance at a shifted affine parameter where 
the shift $\delta\tilde{\lambda}$ is defined by the relation 
\begin{equation}\label{defdeltalambda}
1+{z}(\tilde{\lambda}_s+\delta\tilde{\lambda})=1+\bar{z}(\tilde{\lambda}_s) \,, 
\end{equation}
where $1+\bar{z}(\bar{\eta})=a^{-1}(\bar{\eta})$ is the standard redshift in the RW background spacetime.
These definitions would allow us later to 
substitute the standard RW relation $\tilde{\lambda}_s(\bar{z})$ into Eq. (\ref{dl2}).
Using Eq. (\ref{dl2}) and recalling that $\bar{D}_A=\tilde{\lambda}_s-\tilde{\lambda}_o$ we find that
\begin{equation}\label{dl3}
{D}_L(\tilde{\lambda}_s+\delta \tilde{\lambda})\approx(1+{z}) 
[\bar{D}_A(\tilde{\lambda}_s)(1+\tilde{z})+\delta \tilde{\lambda}+\delta \tilde{D}_A(\tilde{\lambda}_s)]\,.
\end{equation}
We now calculate $\tilde{z}$  and $\delta\tilde{\lambda}$ and substitute their expressions into Eq. (\ref{dl3}).

First, we consider  $\tilde{z}$.  Using the relation $\tilde{u}^\mu\approx(1-\psi,\bold{v})$ together with 
 the metric (\ref{metric}) and the redshift definition (\ref{redshift}) 
applied to the conformal space, we find that
\[
\tilde{z}\approx{\left(\psi+\bold{v}\cdot \bold{n}- \delta \tilde{k}^0\right)}^{\tilde{\lambda}_s}_{\tilde{\lambda}_o}\,.
\]
To calculate the quantity ${( \delta \tilde{k}^0)}^{\tilde{\lambda}_s}_{\tilde{\lambda}_o}$  
we  integrate the null geodesic equation. After integration by parts we obtain  
\begin{equation}\label{finalz}
\tilde{z}\approx{\left(-\psi+\bold{v}\cdot \bold{n}\right)}^{\tilde{\lambda}_s}_{\tilde{\lambda}_o}-
\int_{\tilde{\lambda}_o}^{\tilde{\lambda}_s}(\phi_{,\eta}+\psi_{,\eta})d\tilde{\lambda}\,.
\end{equation}
Next, we consider $\delta\tilde{\lambda}$. 
Combining Eq. (\ref{confredsh}) with Eq. (\ref{defdeltalambda}) we find that 
\begin{equation}
1+\bar{z}[\bar{\eta}(\tilde{\lambda}_s)]\approx\frac{1+\tilde{z}(\tilde{\lambda}_s)} 
{a(\eta_{shift})}\,,
\end{equation}
where
\[
\eta_{shift}=\bar{\eta}(\tilde{\lambda}_s+\delta\tilde{\lambda})+\delta\eta(\tilde{\lambda}_s)\,.
\]
From which we find that
\begin{equation}\label{deltalambda}
\delta\tilde{\lambda}=\delta \eta-\frac{\tilde{z}}{\mathcal{H}_s}\,,
\end{equation}
where we introduced the notation $\mathcal{H}\equiv a_{,\eta}a^{-1}$ and $\delta \eta\equiv\delta x^0$ 
[$\delta x^0$ was defined in  Eq.(\ref{pert1})]. 
Finally, we calculate $\delta \eta $ by 
integrating the null geodesic equation which gives
\begin{equation}\label{deltaeta}
\delta \eta(\tilde{\lambda}_s)=(\phi_o-\psi_o)\bar{D}_A+
\int_{\tilde{\lambda}_o}^{\tilde{\lambda}_s}(\phi_{,\eta}-\psi_{,\eta})(\tilde{\lambda}_s-\tilde{\lambda})d\tilde{\lambda}
+2\int_{\tilde{\lambda}_o}^{\tilde{\lambda}_s}\psi d\tilde{\lambda}+2\int_{\tilde{\lambda}_o}^{\tilde{\lambda}_s}\psi_{,\eta}(\tilde{\lambda}_s-\tilde{\lambda})d\tilde{\lambda}\,.
\end{equation}
To derive this expression  
we used the initial conditions $\delta\eta(\tilde{\lambda}_o)=0$ and $\delta \tilde{k}^0(\tilde{\lambda}_o)=\phi_o+\psi_o$, where 
the last equation is obtained from the requirements 
 $\tilde{k}^\mu \tilde{k}^\nu \tilde{g}_{\mu\nu}=\bar{k}^\mu \bar{k}^\nu \eta_{\mu\nu}=0$  together  
with the initial condition $\delta \bold{\tilde{k}}(\tilde{\lambda}_o)=0 $

\section{Results}\label{results}

We now substitute Eqs. (\ref{bard2},\ref{deltad3})  into  Eq.  (\ref{dl3}) 
and use Eqs.(\ref{finalz},\ref{deltalambda},\ref{deltaeta}) 
after some integrations by parts and a change of notation, $\chi\equiv\tilde{\lambda}$, 
we finally obtain 
\begin{eqnarray}\label{finalresult}\nonumber
&&{D}_L ({z},\bold{n})  \approx(\chi_s-\chi_o)(1+{z})
\Biglb\{1-\psi_s+\bold{v}_s\cdot\bold{n}
 -2\int_{\chi_o}^{\chi_s}\phi_{,\eta}\frac{\chi_s-\chi}{\chi_s-\chi_o}d\chi
+\int_{\chi_o}^{\chi_s}\frac{\psi+\phi}{\chi_s-\chi_o}d\chi\\\nonumber
&&-\frac{1}{(\chi_s-\chi_o)\mathcal{H}_s}\left[ (-\psi+\bold{v}\cdot\bold{n})^{\chi_s}_{\chi_o} 
-\int_{\chi_o}^{\chi_s}(\phi_{,\eta}+\psi_{,\eta})d\chi \right]\\\nonumber
&&-\frac{1}{2}\int_{\chi_o}^{\chi_s} \nabla^2(\phi+\psi)   
\frac{(\chi-\chi_o)(\chi_s-\chi)}{\chi_s-\chi_o}d\chi 
+\frac{1}{2}\int_{\chi_o}^{\chi_s} (\phi_{,\eta\eta}+\psi_{,\eta\eta})   
\frac{(\chi-\chi_o)(\chi_s-\chi)}{\chi_s-\chi_o}d\chi 
 \\
&&+\int_{\chi_o}^{\chi_s}(\phi_{,\eta}-\psi_{,\eta})d\chi 
+\int_{\chi_o}^{\chi_s}(\phi_{,\eta}-\psi_{,\eta})\frac{\chi_o-\chi}{\chi_s-\chi_o}d\chi 
-\frac{1}{2}(\phi_s-\psi_s+\phi_o-\psi_o)\Bigrb\}\,.
\end{eqnarray}
This expression gives 
the luminosity distance as function of the observed redshift and the direction to the source for 
the perturbed RW  metric (\ref{metric}).
Notice that the last three terms in the curly brackets 
vanish for the case of GR where $\phi=\psi$, but may differ from zero for 
modified GR theories.   
Roughly speaking, the  various 
terms in   Eq. (\ref{finalresult}) can be interpreted as representing the following physical phenomena. 
Terms depending on the velocities  represent kinematic Doppler shift arising from the peculiar 
velocities of the observer and the source,
terms depending on the potentials (without derivatives) represent  gravitational redshifts,
and terms depending on a single time derivative of the potentials are analogous to the integrated Sachs-Wolfe (ISW) effect. 
In fact the ISW effect appears in the final term of Eq. (\ref{finalz}).
It has been previously shown \cite{Danieletal} that the ISW effect in modified GR theories 
produces a  modification to the low multipoles of the  CMB anisotropy power spectra.

Consider next the first term in the third line of Eq. (\ref{finalresult}), which 
depends on the Laplacian of the potentials. This term represents gravitational lensing and it 
 agrees with an existing expression for the convergence in the context of modified gravity 
theories \cite{Schmidetal}. 
Consider next the second term in the third line,   
which contains  two time derivatives. 
The fact that this term  is integrated with the standard lensing weight function 
 ${(\chi-\chi_o)(\chi_s-\chi)}/(\chi_s-\chi_o)$
signals that this term is also 
a  lensing term. However, it is smaller than the leading lensing term by 
$\sim (aH/k)^2$, where  $k$ denotes the comoving wave number of a perturbation, and $H$ denotes the Hubble 
rate. 
Therefore, it is subdominant for subhorizon perturbations. 

There is some ambiguity in the  above classification of the various terms in Eq. (\ref{finalresult}), 
and  some of the terms may be classified differently.  For example,  we may use integration  
by parts to replace some of the boundary terms with  terms 
containing integrals over derivatives of the potentials.   
The ambiguity in the classification originates from the fact that 
only the entire luminosity distance is observable, and 
the different individual terms do not correspond to a gauge invariant expression.

As mentioned in the introduction, in many cases Eq. (\ref{finalresult}) can be simplified by 
specializing to subhorizon perturbations. 
Under this approximation, terms that have a time derivative of the potentials 
are smaller by $\sim aH/k$  with respect to the term containing spatial derivatives.  
Furthermore, under this approximation Eq. (\ref{pecvel}) implies that  
the potential $\psi$  is smaller by $\sim aH/k$ with respect to the velocity terms.
While the potential $\phi$ is unconstrained by relation (\ref{pecvel}), 
 it is reasonable to assume that its magnitude is not much larger then the magnitude of $\psi$. 
For this reason we neglect terms that depend on $\phi$ and $\psi$  (with no derivatives). 
The above considerations gives the approximate expression 
\begin{eqnarray}\nonumber
&&{D}_L^{sub} ({z},\bold{n})  \approx(\chi_s-\chi_o)(1+{z})
\Biglb\{1+\bold{v}_s\cdot\bold{n}
-\frac{ (\bold{v}\cdot\bold{n})^{\chi_s}_{\chi_o}  }{(\chi_s-\chi_o)\mathcal{H}_s}
  \\\nonumber
&&-\frac{1}{2}\int_{\chi_o}^{\chi_s} \nabla^2(\phi+\psi)   
\frac{(\chi-\chi_o)(\chi_s-\chi)}{\chi_s-\chi_o}d\chi 
\Bigrb\}\,
\end{eqnarray}
that was discussed in the introduction. 

\acknowledgments 
E. R. would like to thank Rachel Bean for discussion. 
This work was supported by NSF  grants, PHY 0652952, PHY 0457200, PHY 0757735, and 
 PHY 0555216.

\newcommand{\apjl}{Astrophys. J. Lett.}
\newcommand{\aap}{Astron. and Astrophys.}
\newcommand{\cmp}{Commun. Math. Phys.}
\newcommand{\grg}{Gen. Rel. Grav.}
\newcommand{\lr}{Living Reviews in Relativity}
\newcommand{\mnras}{Mon. Not. Roy. Astr. Soc.}
\newcommand{\pr}{Phys. Rev.}
\newcommand{\prsl}{Proc. R. Soc. Lond. A}
\newcommand{\ptrsl}{Phil. Trans. Roy. Soc. London}

\end{document}